\documentclass[12pt]{article}

\usepackage{graphicx}

\usepackage{epsf, times}

\newenvironment{sciabstract}{%
\begin{quote} \bf}
{\end{quote}}

\newcounter{lastnote}

\title{Energy Gaps and Kohn Anomalies \\ in Elemental Superconductors}

\author
{P. Aynajian,$^{1}$ T. Keller,$^{1,2}$ L. Boeri,$^{1}$
\\ S. M. Shapiro,$^{3}$ K. Habicht,$^{4}$ B.
Keimer$^{1,\ast}$\\
\\
\normalsize{$^{1}$Max-Planck-Institute for Solid State Research, Heisenbergstr. 1, D-70569 Stuttgart, Germany}\\
\normalsize{$^{2}$ZWE FRM-II, Technical University of Munich, Lichtenbergstr. 1, D-85748 Garching, Germany}\\
\normalsize{$^{3}$Brookhaven National Laboratory, Upton, New York 11973-5000, USA}\\
\normalsize{$^{4}$Hahn-Meitner-Institute, Glienicker Str. 100, D-14109 Berlin, Germany} \\
\\
\normalsize{$^\ast$To whom correspondence should be addressed;
E-mail:  b.keimer@fkf.mpg.de.}}

\date{}

\begin{document}


\baselineskip24pt


\maketitle


\begin{sciabstract}
The momentum and temperature dependence of the lifetimes of acoustic
phonons in the elemental superconductors Pb and Nb was determined by
resonant spin-echo spectroscopy with neutrons. In both elements, the
superconducting energy gap extracted from these measurements was
found to converge with sharp anomalies originating from
Fermi-surface nesting (Kohn anomalies) at low temperatures. The
results indicate electron many-body correlations beyond the standard
theoretical framework for conventional superconductivity. A possible
mechanism is the interplay between superconductivity and spin- or
charge-density-wave fluctuations, which may induce dynamical nesting
of the Fermi surface.
\end{sciabstract}

\clearpage

Over the past half-century, a comprehensive framework based on the
Bardeen-Cooper-Schrieffer formulation \cite{BCS} has been
developed for the interpretation of experimental data on
superconductors. Although this framework has been challenged by
the discovery of high-temperature superconductivity, it provides a
remarkably successful description of the physical properties of
conventional low-temperature superconductors \cite{Schrieffer}.
Even today, however, the prediction of two of the most important
quantities characterizing a superconductor -- the transition
temperature and the energy gap at the Fermi level -- from first
principles presents a formidable challenge to theory, because they
depend exponentially on material-specific parameters such as the
phononic and electronic densities of states and the
electron-phonon coupling \cite{Gross}. We present neutron
scattering data on the lifetimes of acoustic phonons in Pb and Nb
(the two elements with the highest superconducting transition
temperatures, $T_c = 7.2$ and 9.3 K, respectively) that shed new
light on the energy gap in conventional superconductors.

The energy gap can be directly determined in phonon lifetime
measurements, because electron-phonon scattering is suppressed
(and the phonon lifetimes are thus enhanced) for energies below
the gap. Our data indicate a surprising relationship between the
superconducting gap and the geometry of the Fermi surface, which
also leaves an imprint on the phonon lifetimes \cite{Kohn}: For
phonon wave vectors connecting nearly parallel segments of the
Fermi surface, the electron-phonon scattering probability is
enhanced, and lifetime extrema (termed Kohn anomalies) are
generally expected. We have discovered hitherto unknown Kohn
anomalies in both Pb and Nb, and found that the low-temperature
energy gap coincides with such an anomaly in both materials. This
phenomenon has not been anticipated by the standard theoretical
framework for conventional superconductors.

Both Kohn anomalies \cite{Brockhouse,Stedman,NbDisp,Moncton} and
superconductivity-induced phonon renormalization \cite{Shapiro} have
been observed by inelastic neutron scattering. However, as the
requisite energy resolution is difficult to obtain, these
investigations have been limited to a few selected materials, and
both effects have thus far not been studied accurately in the same
material. The systematic investigation reported here was made
possible by recent advances in resonant spin-echo spectroscopy with
neutrons \cite{Bayrakci,Keller,trisp}, which have enabled the
determination of the lifetimes of dispersive excitations with
$\mu$eV energy resolution over the entire Brillouin zone. Briefly,
the spin echo is generated on a triple-axis spectrometer by using
radio-frequency magnetic fields to manipulate the spin polarization
of neutrons scattered from a crystal before and after the scattering
event. The excitation lifetime is then extracted from the spin-echo
decay profile.

The measurements were taken on high-purity Pb and Nb single
crystals. The resulting spin-echo decay profiles for selected
transverse acoustic phonons in Pb and Nb (Fig. 1) are well described
by exponentials, corresponding to Lorentzian phonon spectral
functions; deviations from Lorentzian lineshapes were not found
within the experimental error. The spin-echo decay rate
(proportional to the phonon linewidth and inversely proportional to
its lifetime) decreases upon lowering the temperature, reflecting
the loss of the electron-phonon decay channel in the superconducting
state. The nonzero decay rate at the lowest temperatures is due to
instrumental limitations, which can be quantitatively determined
based on the phonon dispersion relations and the mosaic spreads of
the single-crystal samples (see SOM \cite{SOM}). The intrinsic
Lorentzian phonon linewidths, $\Gamma$, are extracted by fitting the
decay profiles to exponentials (lines in Fig. 1) and correcting for
this instrumental contribution.

Fig. 2A shows the intrinsic wave vector dependent linewidths of the
lowest-energy, transverse acoustic phonon mode $T_1$ of Pb along
$q=(\xi,\xi,0)$. At all temperatures, sharp anomalies in the phonon
linewidths are seen at $\xi \sim 0.25$, $\sim 0.35$, and $\sim 0.50$
reciprocal lattice units (r.l.u.). Although the phonon spectrum of
Pb has been studied extensively by conventional neutron
spectroscopy, these particular features have not been recognized
because of insufficient energy resolution. Fig. 2B shows that the
same features also appear in the phonon dispersion relation: maxima
in the phonon linewidth coincide with characteristic S-shaped
deviations from the $q$-linear dispersion, as stipulated by the
Kramers-Kronig relation that holds for all excitations in solids.
Artefacts associated with the new measurement method would generally
not be Kramers-Kronig consistent and can thus be ruled out. This
implies that phonons with the anomalous wave vectors shown in Fig.
2A are intrinsically unstable towards decay into other elementary
excitations. In principle, the decay products can be either other
phonons (generated, for instance, by anharmonic terms in the lattice
potential) or electron-hole pairs (originating from Kohn anomalies).
The features at $\xi \sim 0.35$ and 0.5 can be associated with Kohn
anomalies, because these wave vectors are known as nesting vectors
of the Fermi surface \cite{Anderson}. Indeed, Kohn anomalies have
been observed at these wave vectors in the longitudinal phonon
branch of Pb \cite{Brockhouse,Stedman}, (see SOM\cite{SOM}). The
origin of the feature at $\xi \sim 0.25$ is more subtle, because
this wave vector does not match any known spanning vector of the
Fermi surface. A possible origin is a three-phonon decay process
previously observed in the spectrum of phonons in liquid helium,
which are unstable because their phase velocity exceeds the velocity
of sound \cite{HeDisp,3phonon}. Indeed, accurate measurements of the
phonon dispersions in Pb (Fig. 2B) show that the phonon phase
velocity exceeds the sound velocity around $\xi \sim 0.25$,
presumably as a consequence of the dispersion anomaly at $\xi \sim
0.35$. This process has thus far not been observed in solids and
deserves further investigation. Anharmonic terms in the lattice
potential may also contribute to the anomaly.

We will focus on the influence of superconductivity on the phonon
linewidths below $T_c = 7.2$ K. As the superconductor is cooled
below $T_c$, the electron-hole decay channel is closed (and
$\Gamma$ is reduced) below the energy gap $2 \Delta (T)$. This
effect is observed at low wave vectors $\xi$ in Fig. 2A. In
particular, $\Gamma \rightarrow 0$ for $T << T_c$ around $\xi =
0.32$ (corresponding to a phonon energy of 2.47 meV, below the
low-temperature limit of $2\Delta \sim 2.7$ meV known from
tunnelling measurements \cite{Gasparovic}). For lower energies
around $\xi \sim 0.25$, however, $\Gamma$ remains nonzero even at
the lowest temperatures, supporting the notion that the linewidth
anomaly at this wave vector originates from the three-phonon
down-conversion process discussed above and/or lattice
anharmonicity, and not from electron-hole pair production. We have
removed the contribution of this process for clarity and show only
the phonon linewidth $\Gamma_{e-p}$ directly attributable to the
electron-phonon interaction (Fig. 3). As expected, $\Gamma_{e-p}$
exhibits a maximum due to the pileup of electronic density of
states above $2 \Delta (T)$, which moves to progressively higher
energies upon cooling and closely tracks the energy gap determined
in prior tunnelling measurements \cite{Gasparovic} (inset in Fig.
3). Surprisingly, however, the superconductivity-induced maximum
of $\Gamma_{e-p}$ merges with the Kohn anomaly as $T \rightarrow
0$. At $T = 0.5$ K, both anomalies are indistinguishable within
the measurement error.

In order to explore whether the coincidence of $2 \Delta (T
\rightarrow 0)$ and the Kohn anomaly in Pb is accidental, we have
performed similar experiments on phonons in Nb, an elemental
superconductor with a different Fermi surface and phonon spectrum.
Fig. 4A shows the momentum-dependent linewidths of the transverse
acoustic phonon branch along $(\xi,0,0)$ in Nb at temperatures above
and below $T_c=9.3$ K. The data above $T_c$ are in fair agreement
with prior work \cite{Shapiro}, but they reveal several sharp
features that have not been been identified before. Based in part on
ab-initio lattice dynamical calculations, they can be identified as
Kohn anomalies (see below). The existence of a Kohn anomaly  at $\xi
\sim 0.17$ persisting up to room temperature has been suggested
based on prior experimental work \cite{NbDisp,NbAnomaly}. As
described above for Pb, the linewidths are reduced below and
enhanced above the gap for quasiparticle-pair production,
$2\Delta(T)$, in the superconducting state, and the low-temperature
electron-phonon linewidth shows the expected dependence on wave
vector (or energy). Similar to the observation in Pb, the $2\Delta(T
\rightarrow 0)$ extracted from the low-temperature $\Gamma_{e-p}$ of
Nb again coincides with the lowest-energy Kohn anomaly within the
experimental error (Fig. 4B).

To help interpret these observations, we have calculated the phonon
dispersions and linewidths in the framework of ab-initio density
functional perturbation theory in the local-density approximation
(LDA) (see SOM \cite{SOM}), on a very fine mesh of
$\mathbf{q}$-points in reciprocal space. The phonon frequencies were
obtained by diagonalization of the dynamical matrices, and the
electron-phonon linewidths by Allen's
formula~\cite{Allen:linewidths}.
%
%
%
%
The results are in reasonable overall agreement with the
experimental data (Figs. 3 and 4). In particular, both the phonon
frequencies and the linewidths associated with Kohn anomalies in
the high-energy transverse acoustic phonons of Nb (Fig. 4A) and in
the longitudinal phonon of Pb (see SOM~\cite{SOM}) are well
described, indicating that the resolution of the calculations is
sufficient to reproduce subtle structures in $\mathbf{q}$-space.

The lowest-lying Kohn anomalies in the transverse-acoustic phonon
branches of both Pb and Nb are, however, not reproduced by the
calculations (Figs. 3A and 4A,B). These anomalies therefore
originate in factors not included in the calculations, such as the
relativistic spin-orbit coupling, phonon non-adiabaticity
\cite{Pisana}, or many-body correlations beyond Allen's formula
\cite{Cappelluti,Dolgov} or the LDA. As the Kohn anomalies in Pb and
Nb are of comparable strength, the spin-orbit coupling (which is
much stronger in Pb than in Nb) cannot be responsible. Because of
the large Fermi energies of both materials, non-adiabatic
electron-phonon coupling effects should also be extremely weak.

This leaves electron correlation effects beyond the LDA as the most
likely mechanism responsible for the low-energy Kohn anomalies. It
seems reasonable to assume that the same correlations are also
responsible for the observed coincidence of $2\Delta(T \rightarrow
0)$ with the same anomalies. As the anomalies persist to
temperatures above 100 K, superconducting fluctuations are unlikely
to be directly responsible. We note, however, that the formation of
spin or charge density waves driven by electron correlations has
been predicted for Pb and other elemental metals \cite{Overhauser}.
Although extensive searches for static density waves in simple
metals have been unsuccessful, it is conceivable that fluctuations
characteristic of such states dynamically enhance the nesting
properties of the Fermi surface, and hence the propensity for Kohn
anomalies in the phonon spectrum. Indeed, experiments on
charge-density-wave materials such as NbSe$_2$ have revealed Kohn
anomalies \cite{Moncton} and Fermi-surface ``pseudogaps"
\cite{Borisenko,Kiss} in the extended fluctuation regime at
temperatures well above the onset of static density-wave order.
Detailed theoretical work is required to assess whether interference
between density-wave and superconducting correlations can limit the
growth of the superconducting gap and lead to the observed
convergence of both energy scales at low temperatures.

Our experiments on two different elemental superconductors
demonstrate that the low-temperature limit of the superconducting
energy gap coincides with low-lying Kohn anomalies in transverse
acoustic phonons. As both superconductors exhibit different lattice
structures, phonon spectra, Fermi surfaces, and superconducting
gaps, this coincidence cannot be accidental. While its origin is
presently unclear, a specific scenario to explore in future
theoretical work is the interplay between density-wave and
superconducting correlations. It is interesting to point out a
possible analogy to research on high-temperature superconductors,
where an anomalous coincidence of the superconducting gap with a
weakly temperature-dependent ``pseudogap" has recently been reported
in some regions of momentum space \cite{Tanaka,Terashima,Lee}.

\clearpage

\noindent {\bf Fig. 1.} Spin-echo decay profiles of transverse
acoustic phonons at $q=(0.26,0.26,0)$, $E=2.32$ meV, in Pb (upper
two curves), and $q=(0.11,0,0)$, $E=2.06$ meV, in Nb (lower two
curves) at selected temperatures. The spin polarization of the
beam at the detector is plotted versus the spin-echo time $\tau$
\cite{Bayrakci,Keller}. The lines are the results of fits of
exponentials (corresponding to Lorentzian spectral functions) to
the data. The inset shows a conventional triple-axis scan through
the phonon in Pb.

\noindent {\bf Fig. 2.} {\bf (A)} Linewidths of transverse acoustic
phonons along $q=(\xi,\xi,0)$ in Pb at selected temperatures. The
data were obtained by correcting the measured spin-echo decay rates
for instrumental effects (see SOM \cite{SOM}). The grey symbols are
the results of ab-initio lattice-dynamical calculations, as
described in the text. {\bf (B)} Dispersion relation of the same
phonon extracted from triple-axis data. The inset shows the phonon
phase velocity ($E/q$) computed from the data. The blue line in the
main panel and the black line in the inset represent the
experimentally determined sound velocity \cite{elastic}.

\noindent {\bf Fig. 3.} Contribution of the electron-phonon
interaction to the linewidth of the transverse acoustic phonon along
$q=(\xi,\xi,0)$ in Pb. The corresponding phonon energy $E$ is
provided by the scale at the top; note the nonlinear $E$-versus-$q$
relationship (Fig. 2B). The lines are results of least-squares fits
to the Bardeen-Cooper-Schrieffer (BCS) excitation spectrum function
\cite{Dynes}. The inset shows the temperature dependence of the
superconducting energy gap (filled squares) and Kohn anomaly (open
triangles) extracted from the fits. The line in the inset shows the
BCS expression for the superconducting gap \cite{Schrieffer}, which
was experimentally confirmed by tunnelling spectroscopy
\cite{Gasparovic}.

\noindent {\bf Fig. 4.} {\bf (A)} Linewidths of transverse acoustic
phonons along $q=(\xi,0,0)$ in Nb at two different temperatures. The
grey symbols are the results of lattice-dynamical calculations, as
described in the text. {\bf (B)} Blowup of the low-$q$ segment of
panel A. The corresponding phonon energy $E$ is provided by the
scale at the top. The lines are guides-to-the-eye.

\newpage

\begin{figure}[htb]
\begin{center}
\includegraphics[width= \columnwidth, angle=0]{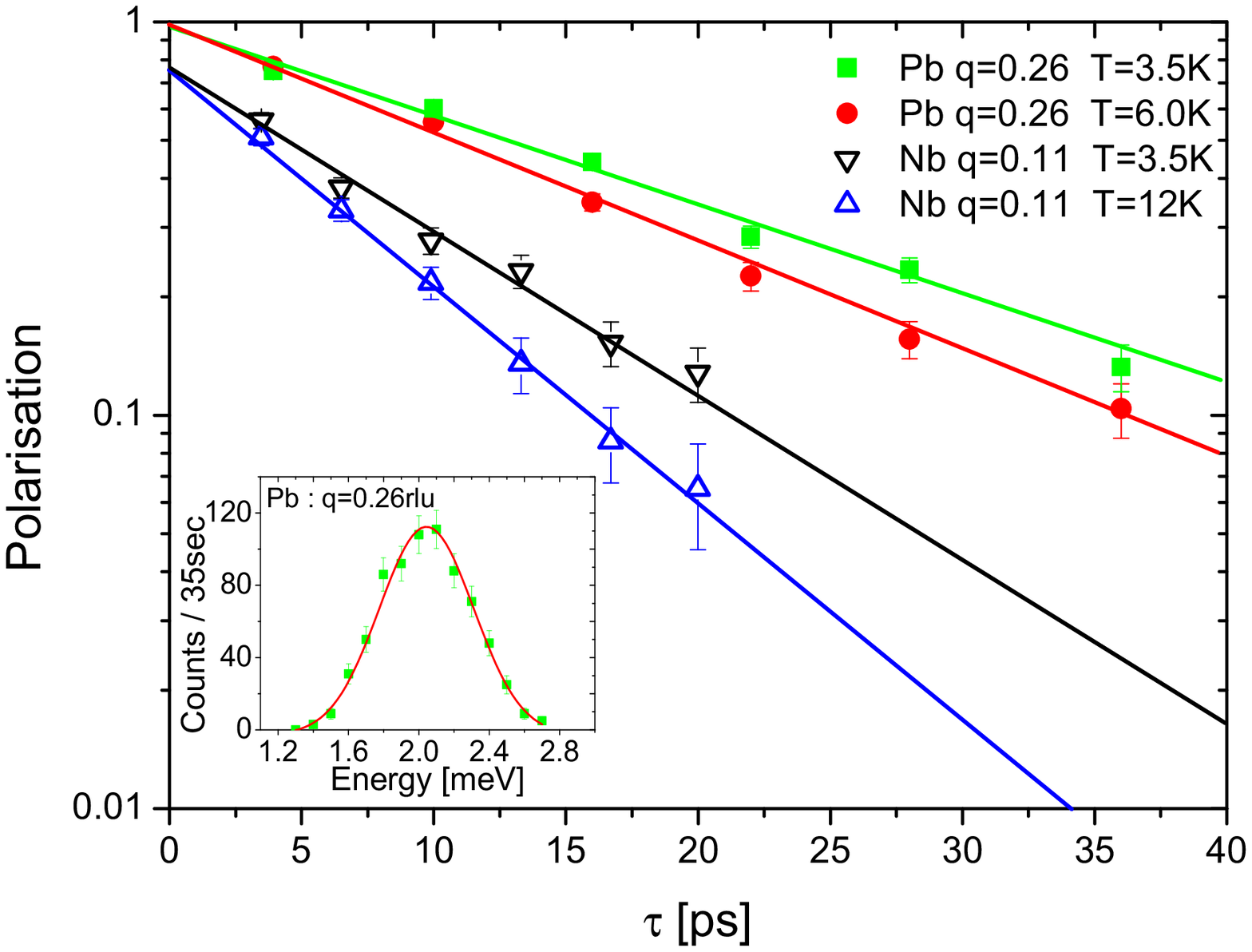}
\caption{
}
\end{center}

\end{figure}

\newpage

\begin{figure}[htb]
\begin{center}
\includegraphics[width= \columnwidth, angle=0]{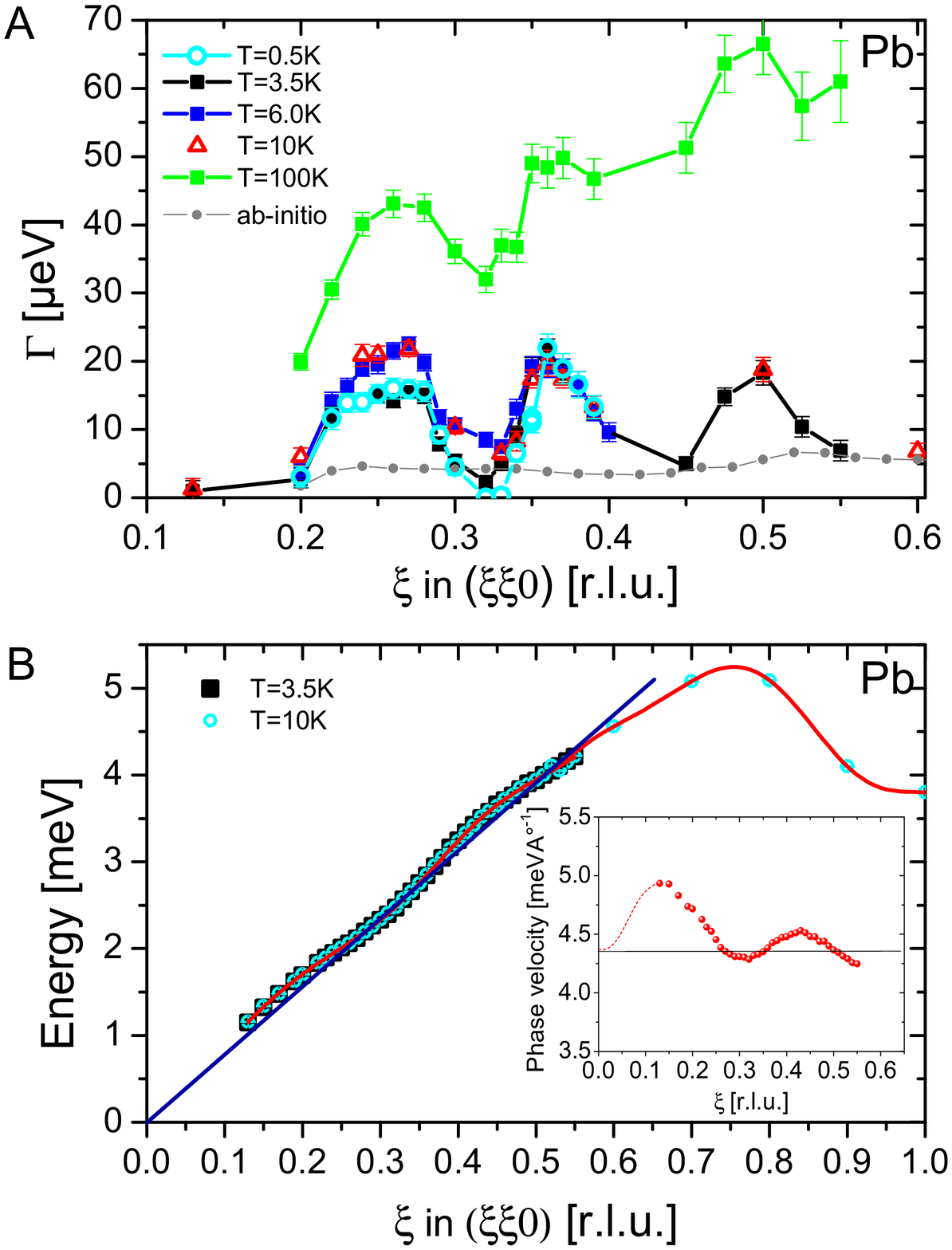}
\end{center}
\caption{
}
\end{figure}

\newpage

\begin{figure}[htb]
\begin{center}
\includegraphics[width= \columnwidth, angle=0]{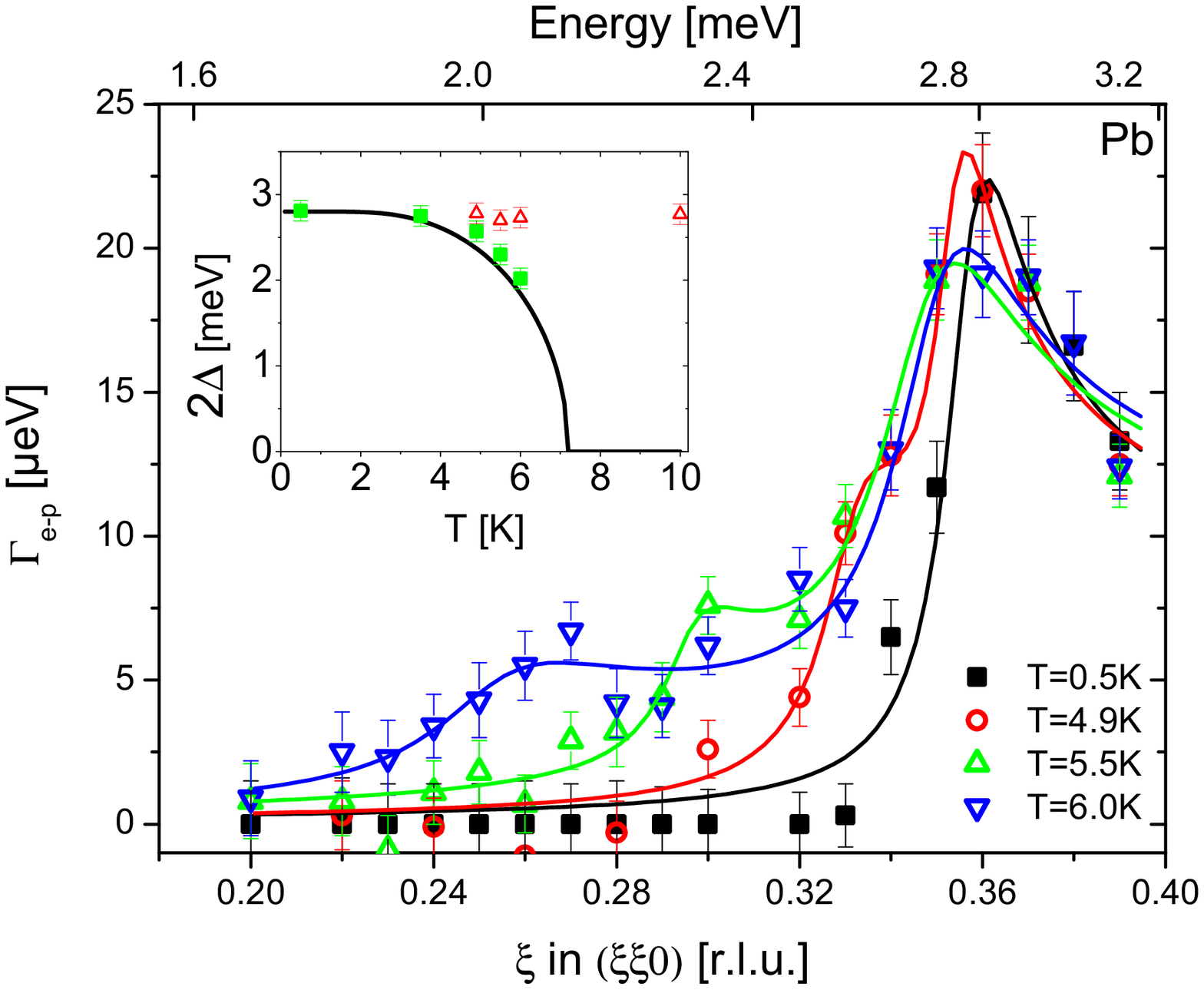}
\end{center}
\caption{
}
\end{figure}

\newpage

\begin{figure}[htb]
\begin{center}
\includegraphics[width= \columnwidth, angle=0]{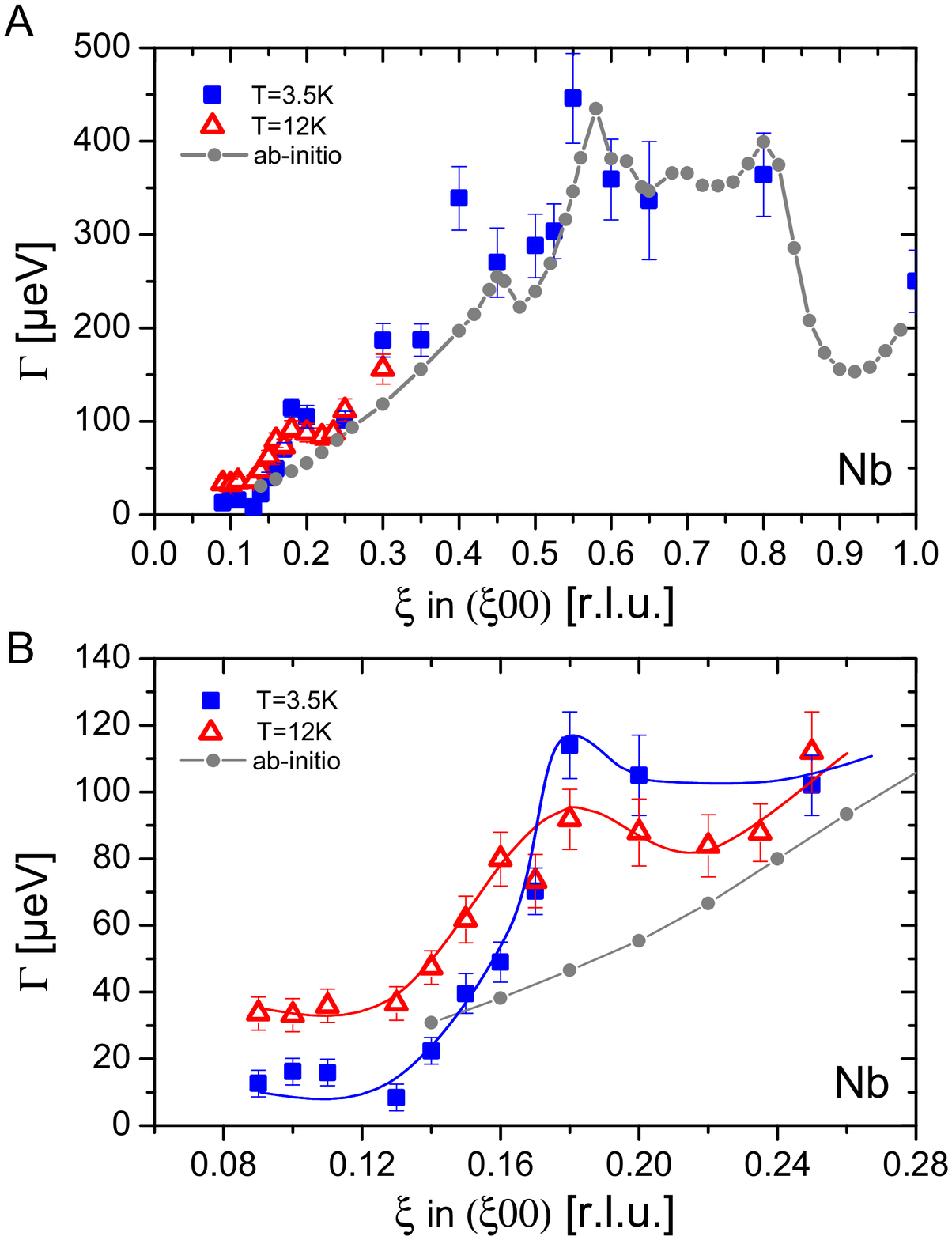}
\end{center}
\caption{
}
\end{figure}
\end{document}